# Improving the efficiency of cascade detection by the Baikal-GVD neutrino telescope

**V.M. Aynutdinov[a], V.A. Allakhverdyan[b], A.D. Avrorin[a], A.V. Avrorin[a], Z. Bardačová[c,d,*], I.A. Belolaptikov[b], E.A. Bondarev[a], I.V. Borina[b], N.M. Budnev[e], V.A. Chadymov[l], A.S. Chepurnov[f], V.Y. Dik[b,g], G.V. Domogatsky[a], A.A. Doroshenko[a], R. Dvornický[c], A.N. Dyachok[e], Zh.-A.M. Dzhilkibaev[a], E. Eckerová[c,d], T.V. Elzhov[b], L. Fajt[d], V.N. Fomin[l], A.R. Gafarov[e], K.V. Golubkov[a], N.S. Gorshkov[b], T. I. Gress[e], K.G. Kebkal[h], I.V. Kharuk[a], E.V. Khramov[b], M.M. Kolbin[b], S.O. Koligaev[i], K.V. Konischev[b], A.V. Korobchenko[b], A.P. Koshechkin[a], V.A. Kozhin[f], M.V. Kruglov[b], V.F. Kulepov[j], Y.E. Lemeshev[e], M.B. Milenin[a,†], R.R. Mirgazov[e], D.V. Naumov[b], A.S. Nikolaev[f], D.P. Petukhov[a], E.N. Pliskovsky[b], M.I. Rozanov[k], E.V. Ryabov[e], G.B. Safronov[a], D. Seitova[b,g], B.A. Shaybonov[b], M.D. Shelepov[a], S.D. Shilkin[a], E.V. Shirokov[f], F. Šimkovic[c,d], A.E. Sirenko[b], A.V. Skurikhin[f], A.G. Solovjev[b], M.N. Sorokovikov[b], I. Štekl[d], A.P. Stromakov[a], O.V. Suvorova[a], V.A. Tabolenko[e], B.B. Ulzutuev[b], Y.V. Yablokova[b], D.N. Zaborov[a], S.I. Zavyalov[b], and D.Y. Zvezdov[b]**

[a] *Institute for Nuclear Research, Russian Academy of Sciences, Moscow, Russia*

[b] *Joint Institute for Nuclear Research, Dubna, Russia*

[c] *Comenius University, Bratislava, Slovakia*

[d] *Czech Technical University in Prague, Institute of Experimental and Applied Physics, Czech Republic*

[e] *Irkutsk State University, Irkutsk, Russia*

[f] *Skobeltsyn Institute of Nuclear Physics, Moscow State University, Moscow, Russia*

[g] *Institute of Nuclear Physics of the Ministry of Energy of the Republic of Kazakhstan, Almaty, Kazakhstan*

[h] *LATENA, St. Petersburg, Russia*

[i] *INFRAD, Dubna, Russia*

[j] *Nizhny Novgorod State Technical University, Nizhny Novgorod, Russia*

[k] *St. Petersburg State Marine Technical University, St. Petersburg, Russia*

[l] *Moscow, free researcher*

*E-mail:* zuzana.bardacova@fmph.uniba.sk

**Speaker*

*†Deceased*

The deployment of the Baikal-GVD deep underwater neutrino telescope is in progress now. About 3500 deep underwater photodetectors (optical modules) arranged into 12 clusters are operating in Lake Baikal. For increasing the efficiency of cascade-like neutrino event detection, the telescope deployment scheme was slightly changed. Namely, the inter-cluster distance was reduced for the newly deployed clusters and additional string of optical modules are added between the clusters. The first inter-cluster string was installed in 2022 and two such strings were installed in 2023. This paper presents a Monte Carlo estimate of the impact of these configuration changes on the cascade detection efficiency as well as technical implementation and results of in-situ tests of the inter-cluster strings.

## 1. Introduction

Since 2016, the construction of the Baikal-GVD neutrino telescope has been continuing in Lake Baikal [1]. In the 2023 configuration, the telescope consists of 3456 optical modules combined into 12 GVD clusters and two experimental strings. One of the priority tasks of the Baikal project is to study the possibilities of increasing the efficiency of the detector based on the experience of its operation and the results obtained on other neutrino telescopes in recent years. The solution of this task, in particular, will create the necessary background for the development of a next-generation neutrino telescope project with an effective volume of 10 cubic kilometers scale. As experiments on neutrino telescopes have shown, 10 $km^3$-scale detector will allow us to move from observing the diffuse flux of astrophysical neutrinos to studying individual neutrino sources. Research is being conducted in the areas of developing a new deep-sea photodetector (optical module) with an increased sensitivity, exploring the possibility of upgrading the data acquisition system based on fiber-optic technology and optimizing the configuration of the telescope's detection system. In this paper, an option of optimizing the telescope configuration is considered, based on the installation of additional, inter-cluster strings in the geometric centers of each three clusters of the detector. The first experimental version of the inter-cluster string (ICS) was installed in Lake Baikal in April 2022. In 2023, two more ICSs were commissioned. The paper presents the results of calculations of the telescope efficiency for the new configuration, the technical implementation, and the first results of in-situ tests of the ICSs.

## 2. Optimization of Baikal-GVD configuration

The Baikal-GVD neutrino telescope is located in the southern part of Lake Baikal. The depth of the lake at the telescope location is 1366 m. Registration of Cherenkov radiation from neutrino interaction products in Baikal-GVD is carried out by optical modules (OM) [2]. A Hamamatsu R7081-100 photoelectronic multiplier tube (PMT) is used as a photosensitive element of the OM. Optical modules are placed on vertical strings anchored at the bottom of the lake and grouped into clusters. The cluster includes a central string and seven strings evenly spaced around a circle with a radius of 60 meters. Each string holds 36 optical modules placed with a step of 15 meters at depths from 750 to 1275 meters.

Optimization of the Baikal-GVD configuration (the distances between the OMs, strings, and clusters) in order to achieve maximum sensitivity to the astrophysical neutrino flux was carried out for the $E^{-2}$ neutrino energy spectrum. Under this condition, the optimal distance between clusters was found to be 300 m. The energy spectrum of astrophysical neutrinos, later measured by the IceCube, showed a higher value of the spectral index [3]. Taking into account the steeper neutrino spectrum, the distance between clusters was reduced in 2022 from 300 m to 250 m and actions were taken to increase the sensitivity of the inter-cluster area of the telescope. From the point of view of technical implementation, the most effective way to increase the sensitivity of the telescope is to install additional inter-cluster strings in the geometric centers of each triplet of the Baikal-GVD clusters (see Fig. 1).

To estimate the effect associated with the installation of ICSs, the response of the detector was simulated in muon and cascade detection modes. The configuration of the ICS was completely the same as the configuration of conventional telescope strings - 36 optical modules located at distances of 15 meters vertically. Monte Carlo simulation shows that for muon events with a track length exceeding the geometrical dimensions of the telescope the effective area increased in proportion to the increase in the number of optical modules. That means that there is no significant effect from the ICS installation. One can only note the more uniform sensitivity of the telescope with ICS to the flux of near-horizontal muons.

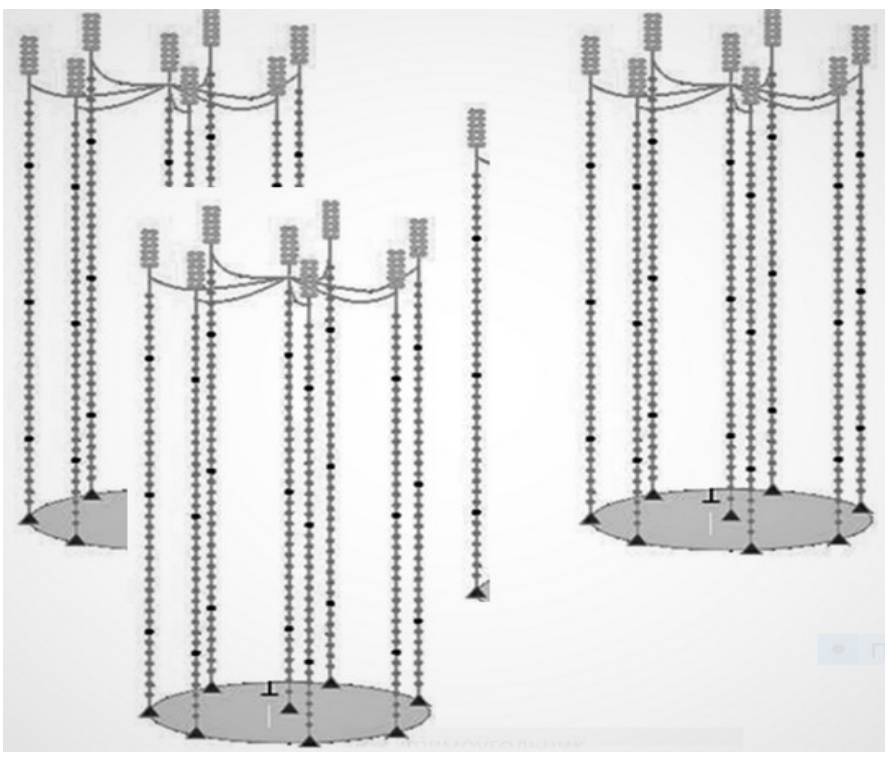

**Figure 1:** Inter-cluster string located in the geometric center of three Baikal-GVD clusters.

The situation is significantly different for cascade events, which are relatively compact light sources. To estimate the effect of the ICS installation, $10^4$ electron neutrino interaction vertices were simulated for the configuration shown in Fig. 1. The azimuthal angle and the cosine of the zenith angle of the cascade were uniformly simulated at each vertex. The primary neutrino energy was uniformly simulated over the $E^{-2.46}$ spectrum from 1 TeV to $10^5$ TeV for each direction. The event selection criterion was a limit on the number of triggered channels more than 30. This requirement ensures reliable suppression of background from atmospheric muon bundles. The event distribution on distance ρ to the intercluster string for the two telescope configurations, respectively (distances between clusters 250 m and 300 m), is shown in Fig. 2.

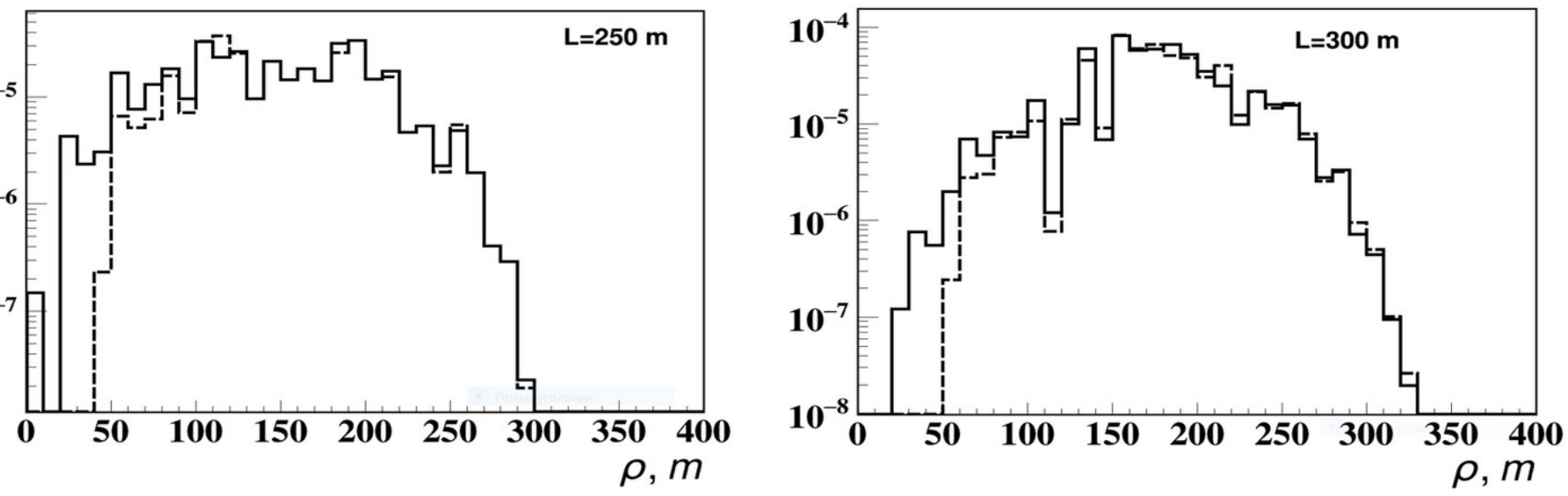

**Figure 2:** Distribution of events on the distance ρ to the geometric center of three clusters for two configurations: the distance between the cluster centers of 250 m (on the left) and 300 m (on the right). Solid lines are the configuration with the ICS, dashed lines are without ICS.

Monte Carlo simulation shows that for a distance between clusters of 250 m, for a configuration consisting of 25 strings (24 strings grouped into 3 clusters and ICS), the number of selected events increased by 10% compared to the conventional configuration without ICS. The increase in the number of events was 5% for the distance between clusters of 300 m. For cascades with an energy more than 100 TeV, where the background from atmospheric muons and neutrinos becomes smaller than the signal from astrophysical neutrinos, the increase in the number of events is 24% for the distance between clusters being 250 m. Thess results show a significant increase in the efficiency of the telescope in a configuration with the ICS.

## 3. Technical implementation of the intercluster string

The configuration of the ICS (see Fig. 3) is basically identical to the conventional Baikal-GVD strings configuration [4]. The ICS consists of three sections of OMs. Each section includes 12 OMs, and a section control module (CM). The OMs are connected to the CM by 92 m long deep-sea cables. The CM controls the OM operation, converts the analog signals of the PMTs into digital form, forms local triggers of the section and time frames of events containing the pulse waveform [5]. The conversion of analog signals is carried out by a 12-channel ADC with a sampling frequency of 200 MHz. The control of the sections operation, formation of the string trigger, and the exchange of the data is provided by a separate deep–sea electronic unit – string control module (SM). The ICS is connected to the cluster control center, just like the conventional cluster strings. Data from the ICS are transmitted to the cluster center via shDSL Ethernet extenders and then transmitted to the shore station via a fiber optic communication line.

ICS is attached to the bottom of the lake by means of an anchor. Its vertical orientation is provided by buoys mounted at the top of the string. Due to the lake currents the string may deviate from the vertical position. The shift of the upper OMs can reach tens of meters. To measure the position of the OMs in real time a positioning system based on acoustic modems (AM) is used [6]. The positioning system of the ICS consists of 4 AMs that provide positioning accuracy of about 0.3 m. AM1 and AM2 are connected to CM1, AM3 and AM4 - to CM3.

At a distance of 270 m from the ICS anchor, a laser calibration light source (laser beacon) connected to CM2 is installed. The power supply (24 V) and control systems (COM Server with RS-485 interface) for the laser and for the AMs are the same, which ensures the unification of all CMs in the string. The laser source emits light at a wavelength of 532 nm, the pulse energy is 0.37 mJ ($\sim 10^{15}$ photons per pulse) with a flash duration of about 1 ns. The laser source includes a light emission system, a radiation stability control system, a controlled attenuator, and a diffuser that ensures the formation of the radiation flux. The attenuator has 6 levels of reduction with the highest attenuation corresponding to a factor of $\sim 10^3$. The laser beacon provides the inter-cluster time calibration, as well as amplitude calibration of the channels, and also allows monitoring the characteristics of the lake water.

Time calibration of the ICS channels is carried out by using LED sources mounted inside each OM. The LED wavelength is 470 nm, and pulse duration is about 5 ns. A light pulse is formed in a $15^{\circ}$cone. Each OM is equipped by two LED sources oriented upwards. Due to narrow radiation cone, emitted light is not detected by the OMs of neighboring strings. Corrections to the ICS time calibration relative to the surrounding clusters are determined using the system of 10 horizontally oriented LEDs mounted on a dedicated support. Typically, few OMs on the string have such instrumentation referred to as LED beacons.

The first ICS was installed and put into operation as part of the Baikal-GVD neutrino telescope in April 2022. Its successful operation throughout the year allowed to continue the installation of the inter-cluster strings – in 2023 two more similar strings were installed. GVD-2023 configuration is shown in Fig. 4. The dots show the strings grouped into clusters. The cluster numbers correspond to the sequence of their commissioning, the asterisks indicate the technological strings with laser calibration sources, circled asterisks show the locations of ICSs that are connected to the clusters 9, 11, and 12.

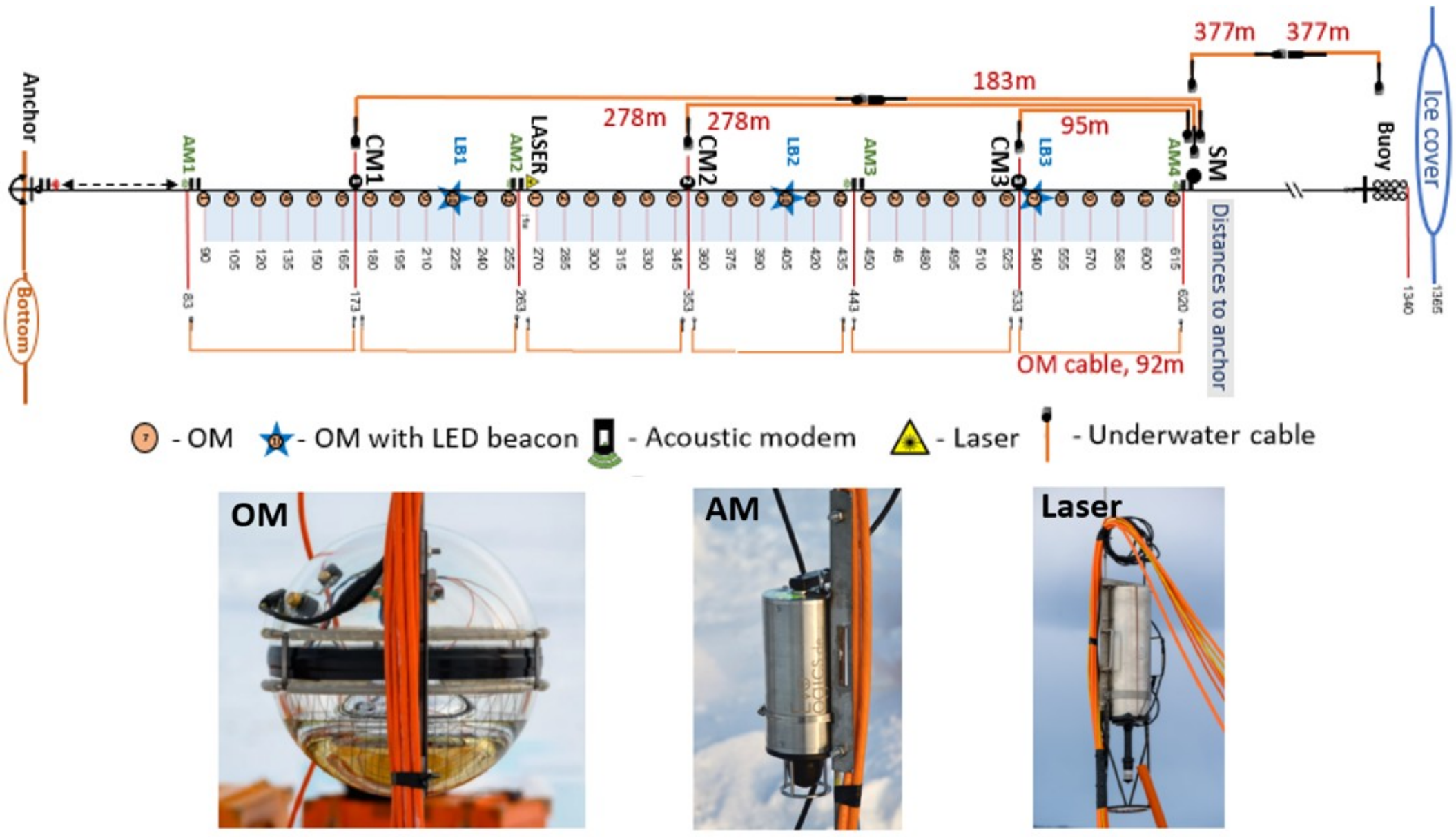

**Figure 3**: Scheme of the mounting and basic elements of the ICS: optical module (OM), acoustic modem (AM) and laser calibration source.

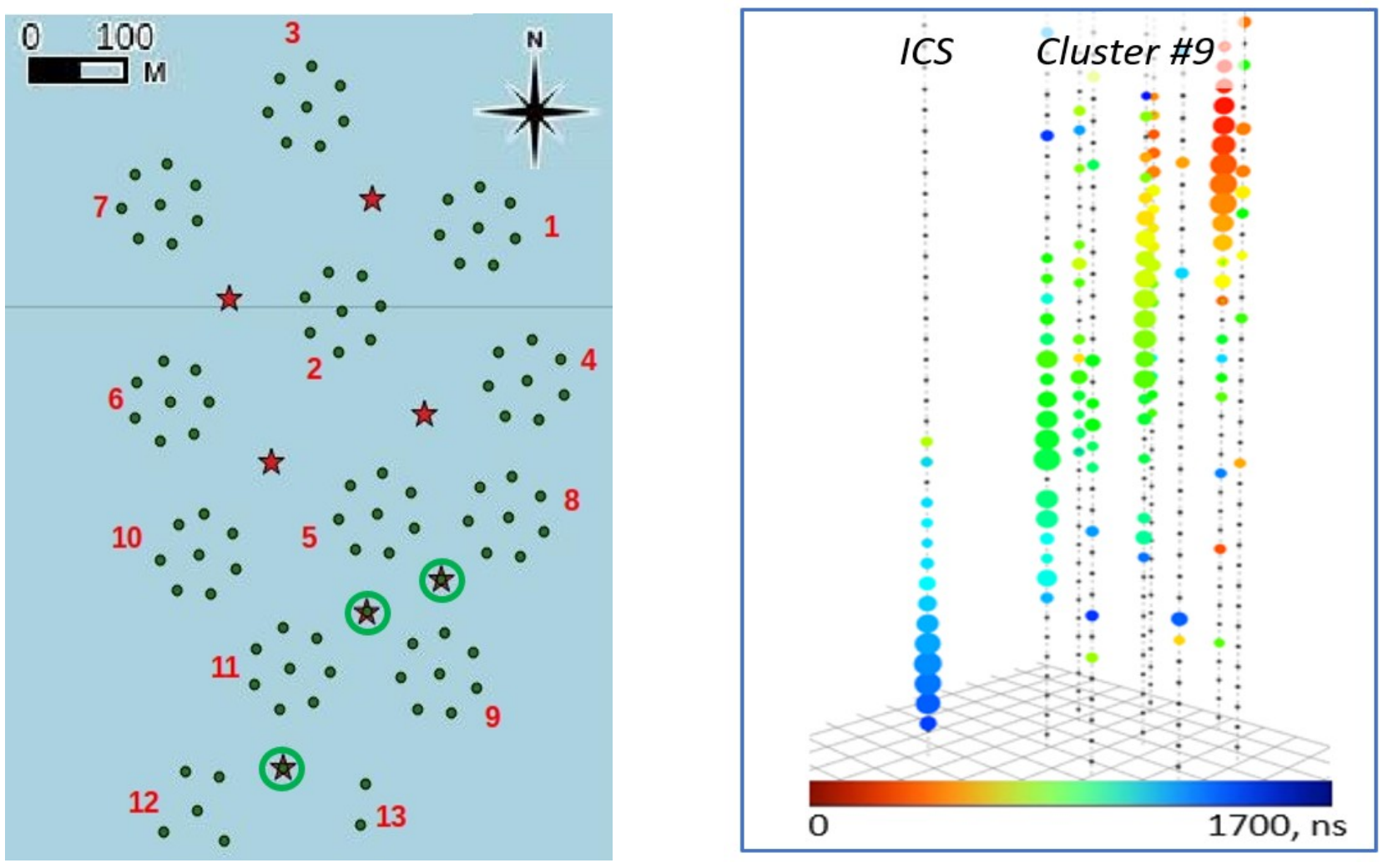

**Figure 4:** Left: Configuration of GVD-2023: asterisks indicate string equipped with lasers, circles show ICSs. Right: An example of muon bundle detection jointly by a GVD cluster and an ICS, the color shows the delays of the signals relative to the first triggered channel.

Analysis of the data sample for 2022-2023 is in progress now. It is planned to estimate the increase of the number of events detected in the cascade channel associated to the ICS installation. In addition to an increase in the number of astrophysical events, an improvement in the background suppression capabilities is also expected. Atmospheric muon bundles are the main background source in the cascade detection mode. Fig. 4 shows an example of such background event. The detection of the muon tracks with the inter-cluster strings provides additional suppression of muon bundles in the telescope.

## 4. In-situ studies of the inter-cluster strings

The accuracy of the event reconstruction in a neutrino telescope depends on the accuracy of PMT pulse time measurement. Time uncertainties are determined by two factors: the accuracy of time offsets of the channels (time calibration) and the uncertainty of signal registration times (time synchronization). The full-scale ICS tests conducted in 2022 included studies of the accuracy of its time calibration and synchronization.

The equipment of the time calibration system was developed for the basic version of Baikal-GVD clusters with the distance between the strings of 60 meters. Under this condition, the accuracy of channel calibration is about 2 ns. With an increase in the distance between the conventional Baikal-GVD strings and ICS (about 80 m), the magnitude of the signal from the calibration LED source decreases, which should affect the accuracy of the time calibration of the ICS. To study this accuracy, several calibration series of measurements were carried out. An example of a calibration event initiated by the LED becon of the ICS and registered on three surrounding clusters is shown in Fig. 5. The triggered channels are highlighted with circles, the color shows the time of the signal registration. The zero-time count is in the middle of the ADC time frame and corresponds to the moment when the cluster trigger is registered by the section module. The adjustment of the time scale is carried out at the stage of setting up the installation using programmable delays of event frames.

The channels within each string were calibrated using LEDs embedded in OMs. Inter-string time calibration of channels was performed using LED beacons. To calculate the relative time offsets between the channels of the ICS and strings of the clusters, the difference ΔT between the time delay expected from the geometry $dT_g$ and measured time delay $dT_m$ between a pair of triggered channels was determined. The flash of the ICS's LED beacon provides several triggered channels in each of the nearby clusters (see Fig. 5). This allows to estimate the accuracy of calibration $\tau$, which is determined by the spread of the measured values of ΔT for different pairs of channels. An example of the time calibration of the ICS relative to the surrounding clusters is shown in Table 1. The table shows the distance R from the ICS to the nearest strings of the surrounding clusters, the average charge Q on the channels in photoelectrons, and the standard deviation of $\tau$ calculated from different pairs of channels.

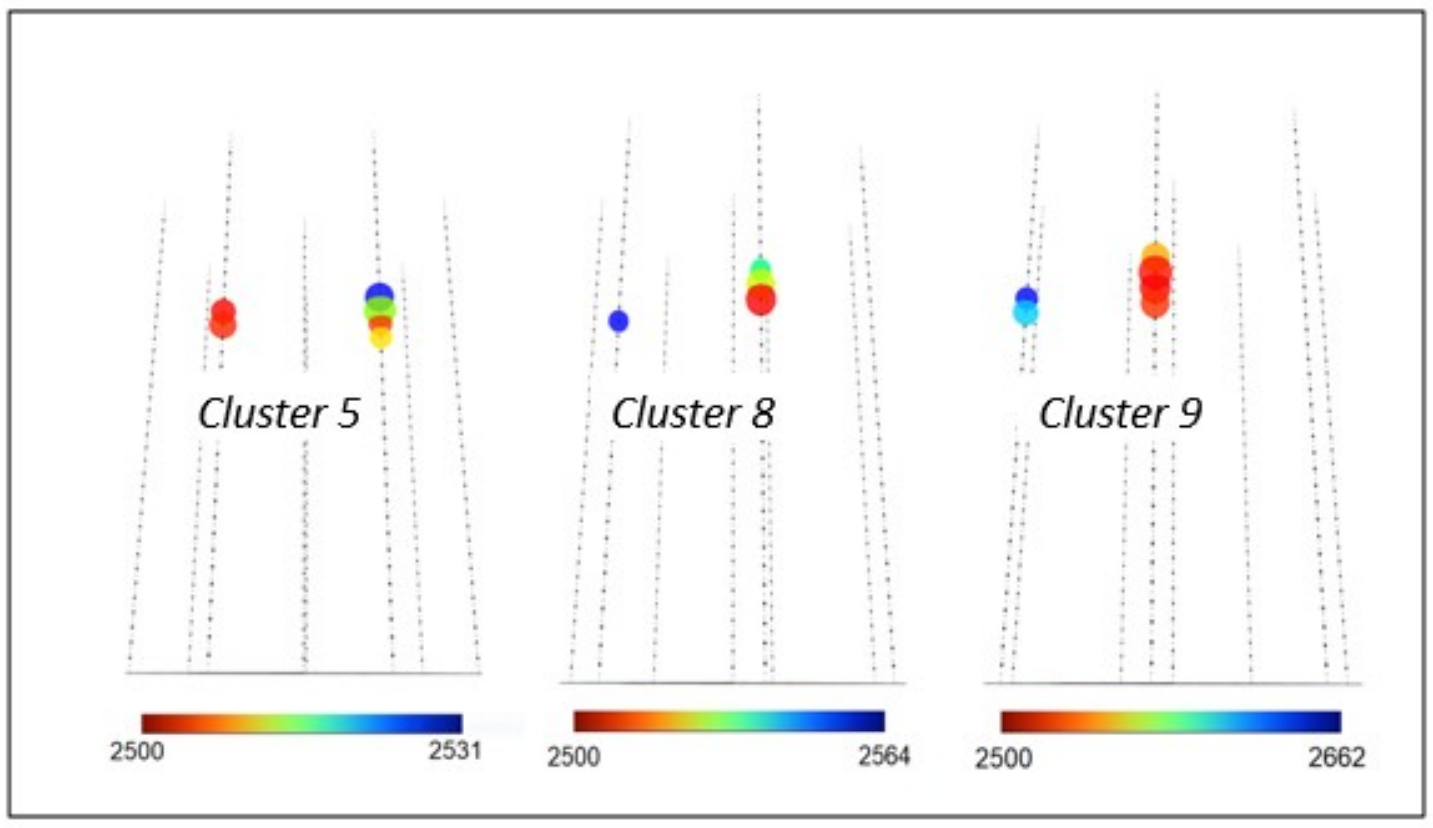

**Figure 5:** View of the calibration event initiated by the ICS LED beacon on the surrounding GVD-clusters in projections to the radiation source. The time scale is presented under each figure in nanoseconds.

**Table 1:** Summary from an LED beacon calibration study of an inter-cluster string relative to nearby clusters (see text).

| Cluster number | *R*, m | *Q*, p.e. | $\tau$, ns |
|---|---|---|---|
| Cluster 5 | 85 | 3.6 | 2.8 |
| Cluster 8 | 80 | 4.6 | 0.4 |
| Cluster 9 | 77 | 3.5 | 2.5 |

The charges of the signals on the calibration channels are about 4 p.e. on average. With such charge values, the uncertainty of the time calibration of the ICS is about 2.5 ns, that is close to the calibration accuracy of the conventional Baikal-GVD strings (2 ns). Such an accuracy is acceptable from the point of view of physics event reconstruction [7]. It should be noted that the measured charge of the triggered channels does not correlate with the distance from ICS to the clusters. This is due to the fact that the amplitudes of the light pulses for different instances of LEDs can vary significantly, which violates the isotropy of the light flux from the LED beacon as a whole.

The Baikal-GVD time synchronization system ensures the operation of all telescope channels in a single time scale. It includes two subsystems that ensure synchronization of channels within one cluster, and synchronization of clusters with each other [7]. The operation of these subsystems is based on different principles. Synchronization of channels within a cluster is carried out using a common trigger generated in the cluster control center and broadcast to all its sections. For inter-cluster synchronization, the time of common trigger is measured in each of them. To study the accuracy of ICS synchronization with clusters, calibration series in the regime of simultaneous illumination of the ICS and clusters by a laser were analyzed. As a parameter characterizing the accuracy of synchronization, the value of the standard deviation (RMS) of the delay of the response times of pairs of synchronized channels (dt) was used. Channels with charges exceeding 10 photoelectrons were selected for analysis. Fig. 6 illustrates the accuracy of the synchronization of the ICS with the surrounding clusters 5, 8, and 9. It should be emphasized that ICS synchronization with cluster 9, of which it is part as the 9th string, was carried out using the common trigger of the cluster, while for clusters 5 and 8, the time between cluster triggers were measured using the "White Rabbit" synchronization system. The accuracy of the synchronization of the ICS with the surrounding clusters was 2.1 – 2.2 ns, which is in a good agreement with the expected value of 2.0 ns, which is determined by the time clock discretization of 5 ns.

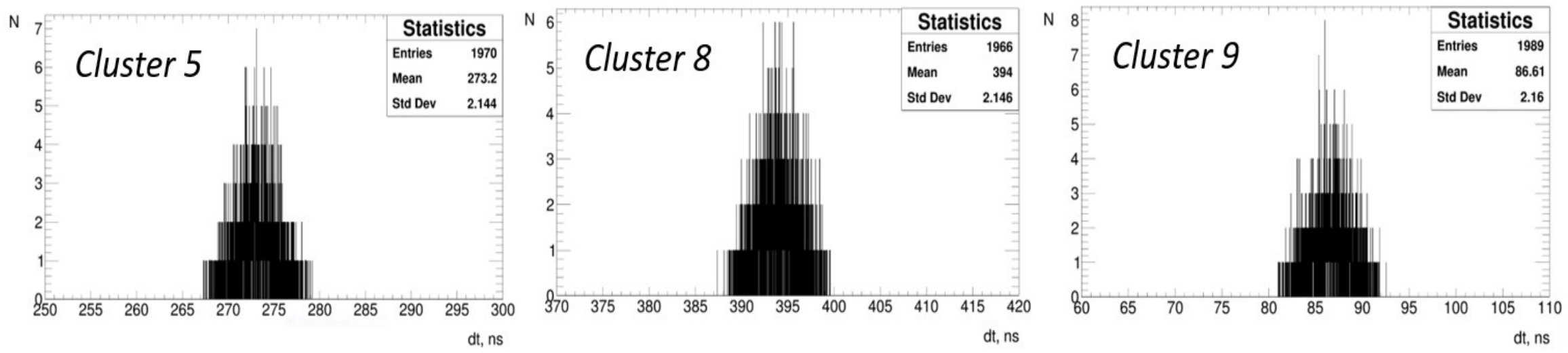

**Figure 6:** Distribution of events on the time difference dt, measured between ICS and clusters 5, 8, and 9.

## 5. Conclusion

Monte Carlo calculations show a significant increase in the efficiency of cascade detection by the Baikal-GVD telescope when inter-cluster strings are added between the clusters. For a configuration of three clusters, the installation of an inter-cluster string provides an increase in the number of events in cascade mode by 10% and 24% for cascade energy above 1 TeV and 100 TeV, respectively. The technical implementation of the inter-cluster strings installation is straightforward. The data acquisition system, deep-sea cable infrastructure, and power supply system of the Baikal-GVD cluster can be easily adapted to serve 9 strings (including ICS) instead of 8. In-situ tests of the ICS showed the correctness and reliability of the equipment operation, and a sufficiently high accuracy of its time calibration (~ 2.5 ns). The accuracy of time synchronization of the ICS was ~2 ns, coinciding with the accuracy of synchronization of the conventional Baikal-GVD strings. Based on the positive experience of operating the first ICS in 2022, two more such strings were installed in 2023. In the future, it is planned to equip all Baikal-GVD clusters with the inter-cluster strings.